\documentclass[10pt,conference]{IEEEtran}
\usepackage{cite}
\usepackage{amsmath,amssymb,amsfonts}
\usepackage{algorithmic}
\usepackage{graphicx}
\usepackage{textcomp}
\usepackage{xcolor}
\usepackage[hyphens]{url}
\usepackage{mathptmx} 
\usepackage{fancyhdr}
\usepackage[normalem]{ulem}
\usepackage[hyphens]{url}
\usepackage{xcolor}
\usepackage{soul}
\usepackage[braket, qm]{qcircuit}
\usepackage{physics}
\usepackage{graphicx}
\usepackage{enumitem}
\usepackage{multirow}
\usepackage{subcaption}
\usepackage{algorithm}
\usepackage{booktabs}
\usepackage{braket}
\usepackage{textcomp}
\usepackage{flushend}
\makeatletter 
\newcommand{\linebreakand}{%
  \end{@IEEEauthorhalign}
  \hfill\mbox{}\par
  \mbox{}\hfill\begin{@IEEEauthorhalign}
}
\makeatother 
\title{Hardware Architecture for a\\ Quantum Computer Trusted Execution Environment}
\author{
\IEEEauthorblockN{Theodoros Trochatos}
\IEEEauthorblockA{
\textit{Electrical Engineering} \\
\textit{Yale University}\\
New Haven, CT, USA \\
theodoros.trochatos@yale.edu
}
\and
\IEEEauthorblockN{Chuanqi Xu}
\IEEEauthorblockA{
\textit{Electrical Engineering} \\
\textit{Yale University}\\
New Haven, CT, USA \\
chuanqi.xu@yale.edu
}
\and
\IEEEauthorblockN{Sanjay Deshpande}
\IEEEauthorblockA{
\textit{Electrical Engineering} \\
\textit{Yale University}\\
New Haven, CT, USA \\
sanjay.deshpande@yale.edu
}
\linebreakand %
\IEEEauthorblockN{Yao Lu}
\IEEEauthorblockA{
\textit{Applied Physics} \\
\textit{Yale University}\\
New Haven, CT, USA \\
physics.lu@yale.edu
}
\and
\IEEEauthorblockN{Yongshan Ding}
\IEEEauthorblockA{
\textit{Computer Science} \\
\textit{Yale University}\\
New Haven, CT, USA \\
yongshan.ding@yale.edu
}
\and
\IEEEauthorblockN{Jakub Szefer}
\IEEEauthorblockA{
\textit{Electrical Engineering} \\
\textit{Yale University}\\
New Haven, CT, USA \\
jakub.szefer@yale.edu
}
}
\begin{document}
\maketitle
\thispagestyle{plain}
\pagestyle{plain}
\begin{abstract}
  The cloud-based environments in which today's and future quantum computers will operate, raise concerns about the security and privacy of user's intellectual property. Quantum circuits submitted to cloud-based quantum computer providers represent sensitive or proprietary algorithms developed by users that need protection. Further, input data is hard-coded into the circuits, and leakage of the circuits can expose users' data. To help protect users' circuits and data from possibly malicious quantum computer cloud providers, this work presented the first hardware architecture for a trusted execution environment for quantum computers. To protect the user's circuits and data, the quantum computer control pulses are obfuscated with decoy control pulses. While digital data can be encrypted, analog control pulses cannot and this paper proposed the novel decoy pulse approach to obfuscate the analog control pulses. The proposed decoy pulses can easily be added to the software by users. Meanwhile, the hardware components of the architecture proposed in this paper take care of eliminating, i.e. attenuating, the decoy pulses inside the superconducting quantum computer's dilution refrigerator before they reach the qubits. The hardware architecture also contains tamper-resistant features to protect the trusted hardware and users' information. The work leverages a new metric of variational distance to analyze the impact and scalability of hardware protection. The variational distance of the circuits protected with our scheme, compared to unprotected circuits, is in the range of only $0.16$ to $0.26$. This work demonstrates that protection from possibly malicious cloud providers is feasible and all the hardware components needed for the proposed architecture are available today.
\end{abstract}
\section{Introduction}
Noisy Intermediate-Scale Quantum (NISQ) quantum computers are being rapidly developed, with machines over $100$ qubits available today~\cite{chow2021ibm} and the industry projects $4000$-qubit or larger devices before the end of the decade~\cite{4000qubits}. Many different types of quantum computers exist, with superconducting qubit quantum computers being one of the types available today to researchers and the public through cloud-based services. The superconducting qubit machines are developed by numerous companies, such as IBM~\cite{ibmquantum}, Rigetti~\cite{rigetti}, or Quantum Circuits, Inc.~\cite{quantumcircuitsinc}. These machines implement quantum computing with superconducting electronic circuits which are operated at about $20$ $mK$ temperatures, by placing the quantum chips in dilution~refrigerators.
Due to the expensive nature of quantum computing equipment, quantum computers are currently available as cloud-based systems. Cloud-based services such as IBM Quantum~\cite{ibmquantum}, Amazon Braket~\cite{amazonbracket}, and Azure Quantum~\cite{azurequantum} already provide access to the NISQ quantum computers remotely for users, for example. In the cloud setting, the cloud provider has full control over the quantum computers. Especially, they can have access to the circuits (and the resulting control pulses) that execute on the quantum computer. Given knowledge of the circuits (or equivalently of the control pulses, which can be reverse-engineered into the circuits), the cloud provider has full access to what the users are executing.
To help protect from the untrusted quantum computer cloud providers, a number of researchers have so far focused on developing delegated quantum computation (DQC), blind quantum computation (BQC), and similar ideas~\cite{childs,universal_blind, aharonov2008interactive, morimae2011ground, Dunjko_20121, Morimae_2012, Morimae_2013, Fitzsimons_20171, Morimae_20121, PhysRevA.87.060301, morimae2013composable, Giovannetti_2013, Mantri_2013, morimae2014verification}.
Most of the work remains theoretical, due to the fact that DQC and BQC require 
a local, trusted quantum computer, as well as, quantum networking to connect the cloud-based and local quantum~computers. 
As an alternative approach that could be deployed today to protect users' quantum circuits and data, this paper proposes a new hardware architecture for the protection of quantum circuits from honest-but-curious quantum computer cloud providers. By leveraging trusted hardware, the proposed Quantum Computer Trusted Execution Environment (QC-TEE).
The goal of the QC-TEE is to protect the users' circuits, and any data embedded in the circuits, from being reverse engineered by honest-but-curious quantum computer~providers.
The key insight behind QC-TEE is that, even if all the digital information is protected with quantum-safe cryptography~\cite{wang2022quantum}, the quantum computing cloud provider has access to the analog control pulses sent between the controller electronics and the dilution refrigerator. These analog pulses cannot be encrypted using digital encryption.
From the control pulses, the provider could reverse-engineer the quantum gates, i.e. quantum operations, and thus circuits, breaking the security and confidentiality of the circuits and any data embedded in the circuits.
If the control pulses can be obfuscated and randomized from the view of the cloud provider, he or she will no longer be able to learn what the user's circuits or data~are.
Based on this insight, we have designed and evaluated the QC-TEE architecture presented in this paper. The main feature of the architecture is the software addition of the decoy pulses on the user's end, and hardware RF switches inside the dilution refrigerator on the quantum computer end to attenuate the decoy control. The operation of the switches and protection of the user's circuits is managed by a simple, trusted {\em hardware security manager}. Both the security manager and switches require very low area and consume minimal power, and could be deployed today.
\subsection{Contributions}
The contribution of this work is the design of the first architecture for trusted execution on cloud-based quantum computers, assuming honest-but-curious cloud providers:
\begin{itemize}[noitemsep]
    \item Novel design of decoy control pulses used to hide the true quantum gates executing on the quantum computer.
    \item Use of simple RF switches inside the quantum computer refrigerator to attenuate the decoy control pulses.
    \item Design of {\em hardware security manager} to control RF switches, with support for various {\em obfuscation levels} and {\em randomized output protection}.
    \item Development of tamper-resistant features to detect any breaches of the trusted boundary formed by the dilution refrigerator.
\end{itemize}
\noindent In addition, the work is evaluated on real IBM quantum computers by executing circuits simulating perfect and imperfect RF switches, and analyzing the influence of imperfect RF switches on circuits through pulses with varied amplitude.
\begin{table}[t]
\caption{\small Comparison to blind quantum execution schemes.}
\label{tab_blind_computation_comparison}
\small
\centering
\begin{tabular}{ |c||p{1.2cm}|p{1.2cm}|p{1.2cm}|p{1.2cm}| }
 \hline
 \textbf{Scheme} & \textbf{Quantum Server} & \textbf{Quantum User} & \textbf{Quantum Network} & \textbf{Trusted Hardware} \\
 \hline
 BQC~\cite{https://doi.org/10.48550/arxiv.1611.10107}  & yes  & yes &  yes & no \\
 DQC~\cite{Dunjko_2014},~\cite{MauRen11} & yes  & yes &  yes & no \\
 QC-TEE (our)   & yes  & no &  no & yes \\
 \hline
\end{tabular}
\end{table}

\section{Background}
\label{background}
This section provides a brief background on quantum computation and the current cloud-based quantum computers.
\subsection{Qubits and Quantum Gates}
A qubit is the fundamental unit of information on a quantum computer and may be represented using a vector $|\psi\rangle = \alpha |0\rangle + \beta |1\rangle$, which is a superposition of the basis states $|0\rangle$ and $|1\rangle$ such that $|\alpha|^2 + |\beta|^2 = 1$. When measured, a qubit collapses to a binary $0$ or $1$ with probabilities $|\alpha|^2$ and $|\beta|^2$, respectively. Similarly, an $n$-qubit system exists in a superposition of $2^n$ basis states and can produce any of the $2^n$ bitstrings depending on the probabilities associated with them upon measurement. Quantum gates such as single-qubit (e.g., Pauli-X gate) or multiple-qubit (e.g., 2-qubit CNOT gate) gates modulate the state of qubits and thus perform computations.
\subsection{Control Pulses}
Microwave pulses are typically used to control superconducting qubits. To operate each native gate on a quantum computer, the necessary control pulses for each gate must be created and supplied to the quantum computer.
The pulses for all native gates on IBM Quantum are preset and their parameters for defining the gates are routinely updated through calibrations to maintain excellent fidelity over time. Pulse parameters are measured and calibrated automatically and are ready to be utilized to create control pulses for quantum~circuits.
\subsection{Cloud-based Quantum Computers}
\begin{figure}[t]
     \centering
         \includegraphics[width=1\linewidth,trim={0cm 0cm 0cm 0cm},clip]{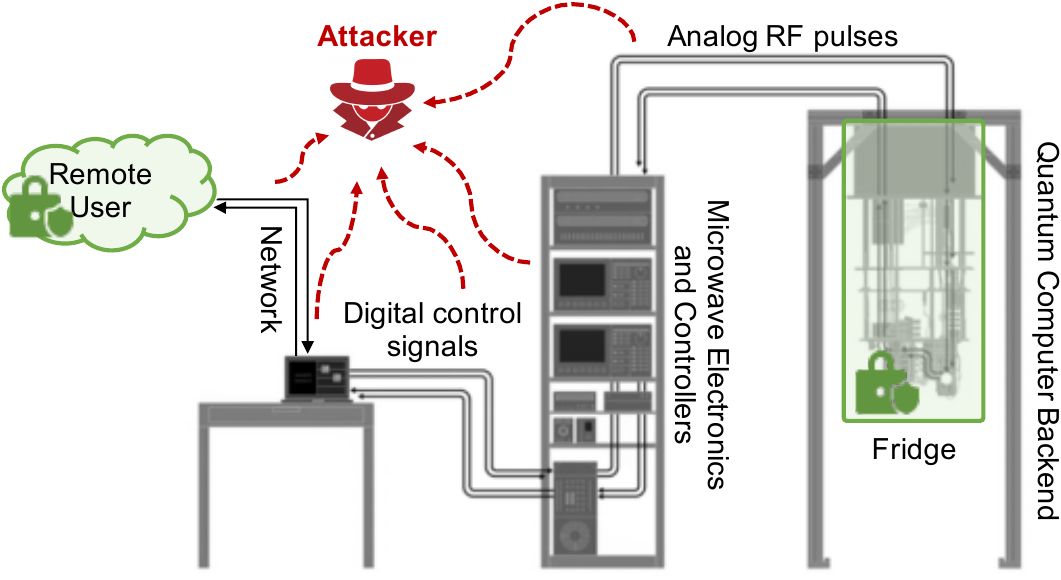}
        \caption{\small Schematic of a typical superconducting quantum computer showing an honest-but-curious cloud provider attempting to spy on the control pulses and equipment outside of the dilution refrigerator. The green boundaries are trusted components.}
        \label{fig_threat_model}
\end{figure}

Most of the cloud-based quantum computer deployments available today follow a similar workflow: users submit jobs, i.e. quantum circuits also called quantum programs, to the cloud provider, then jobs are queued for execution, when the user's job reaches the head of the queue it is executed on the real quantum computer and measurements are returned to the~user.
The remote user is able to develop their code in a high-level language, such as Python. The code is then transpiled into native quantum gates, which can be executed by the target quantum computer, also called ``backend''. The transpiled circuit is then sent to the quantum computer provider as a job that is to be executed on the backend. Remote users can also specify their own custom control pulses. Both native basis gates and custom pulses are supported by our security~architecture.
Based on the request to execute the provided quantum circuit, the quantum computer manager is responsible to schedule the job on the quantum computer, when it is available. Further, the transpiled circuit is converted into control RF pulses, and the controller FPGAs and arbitrary waveform generators (AWGs) are instructed to generate the RF pulses actually sent to the quantum computer hardware. 
The readout of the results, also called ``measurement'', is likewise done by issuing a number of RF control pulses and reading out the response. During measurement operation, the qubit state is collapsed into the classical state and the classical bits are sent back to the user as the computation results. Each job typically consists of thousands of ``shots''. The users collect measurement statistics of all the shots to compute the probabilities of the different classical bit results and typically find the result with the highest probability. The setup of a typical superconducting quantum computer is shown in Figure~\ref{fig_threat_model}.

\section{Threat Model and Assumptions}
\label{threat_model}
\subsection{Threat Model}
\begin{figure*}[t]
     \centering
         \includegraphics[width=0.95\linewidth,trim={0cm 0cm 0cm 0cm},clip]{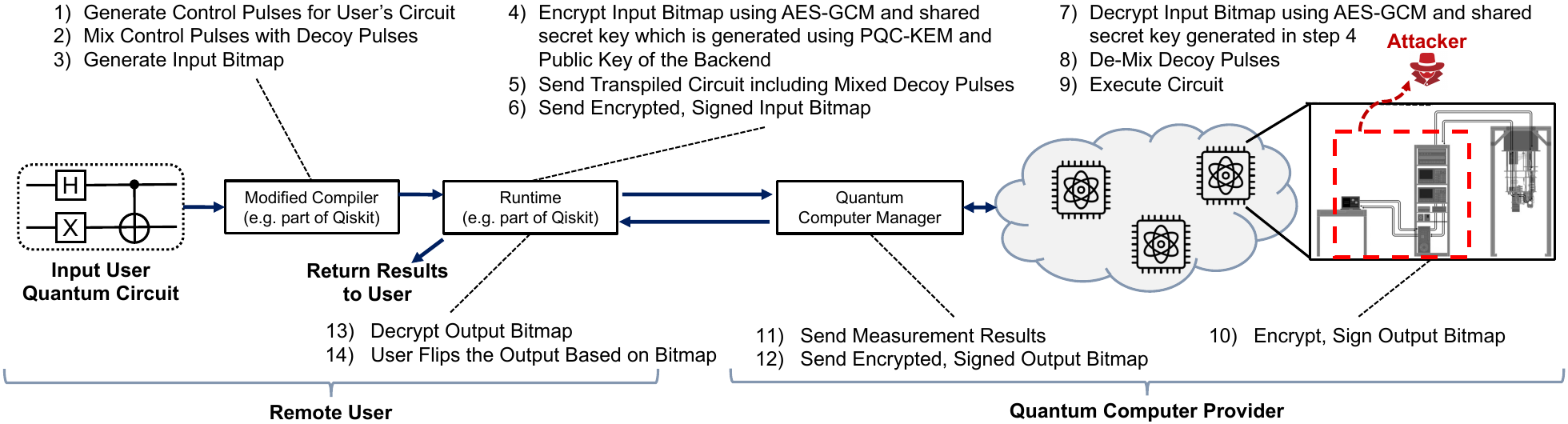}
        \caption{\small Workflow of QC-TEE overlaid on typical cloud-based quantum computation workflow.}
        \label{fig_qctee_workflow}
\end{figure*}
Our work mainly aims to protect users from honest-but-curious cloud providers who want to learn what quantum programs, i.e., quantum circuits, are being executed by the user. We assume the cloud provider has access to all the information and electronics, except the inside of the dilution refrigerator. We assume the honest-but-curious provider can observe any communication (digital and analog), but does not tamper with it. Thus, our goal is to protect from passive attacks, such as information leakage. Active attacks such as fault injection are orthogonal and future work. 
We highlight that the user's circuits are only unobfuscated on the user side and in the fridge, and only the user and the trusted hardware in the fridge have access to the unencrypted information about our decoy pulses. As a result, our scheme also implicitly protects the circuits during transmission on the network between the user and the cloud, or other locations between the user and the fridge, considering there can be passive attacks in these locations.
Similar to classical the CPU package forming trust boundary for security architectures such as Intel SGX~\cite{intelsgx}, we assume the dilution refrigerator encapsulating the quantum computing chips is the trust boundary. Our trusted hardware components are within this boundary. While the refrigerator is physically large, opening and manipulating the refrigerator can be easily detected since it is a manual process accompanied by drastic temperature and pressure changes. Our security architecture includes tamper-detection features, similar to tamper-detection mechanisms available in classical hardware security modules (HSMs)~\cite{mavrovouniotis2013hardware}. Trusted hardware is considered free of bugs or hardware trojans.
\subsection{Assumptions}
On the hardware end, we assume the backend can be modified with our minimal {\em hardware security manager}, which includes classical encryption and decryption engines, a classical true random number generator, and RF switches (details of the hardware components are discussed in the Section~\ref{tee}). We assume the modifications to the backend's hardware are within the cooling power of today's fridges; we do not introduce any complex electronics such as signal generators within the fridge. Further, we assume the manager logic is embedded with a public-private cryptographic key pair and that users can use certificates to authenticate the correct public key corresponding to the target backend. Due to the trusted boundary of the refrigerator, the provider cannot access the keys embedded inside. We assume the added classical hardware logic has been validated or verified and side-channel attacks on this classical hardware logic are out of the scope of this work.
On the software end, following Figure~\ref{fig_qctee_workflow}, we assume the user is able to transpile his or her circuits locally, as it is possible today with Qiskit~\cite{cross2018ibm} or similar frameworks. We assume the users' local environment is secure so the operations required by the scheme can be performed securely without leaking any information. We assume the cloud provider shares correct basis gates information and calibration data with the user and thus the user can transpile his or her circuits effectively. The transpiled circuits can contain any valid gates from the basis library, custom control pulses if the user develops their own control pulses, as well as our decoy gates added by the user, as explained later. In addition to the transpiled circuits, we assume the user is able to send additional classical data, which is the encrypted {\em input bitmap} used to specify which are the decoy control pulses to be attenuated at the quantum computer backend. We assume the user has access to and validates the cryptographic keys corresponding to the backend's hardware and that quantum-safe cryptography is used everywhere classical data is involved.
Note that the honest-but-curious attacker can spy on digital control information between the control computer and control logic, or can spy on the RF control pulses between the control logic and quantum computer, as shown in Figure~\ref{fig_threat_model}. Digital control information and the RF control pulses both effectively contain the same information. The digital control information can easily be encrypted (between the control computer and control logic). However, classical encryption cannot be applied to analog RF pulses. The proposed QC-TEE addresses exactly the problem of protection from an honest-but-curious cloud provider who tries to spy on the control pulses.

\begin{figure*}[t]
     \centering
         \includegraphics[width=1\linewidth,trim={0cm 0cm 0cm 0cm},clip]{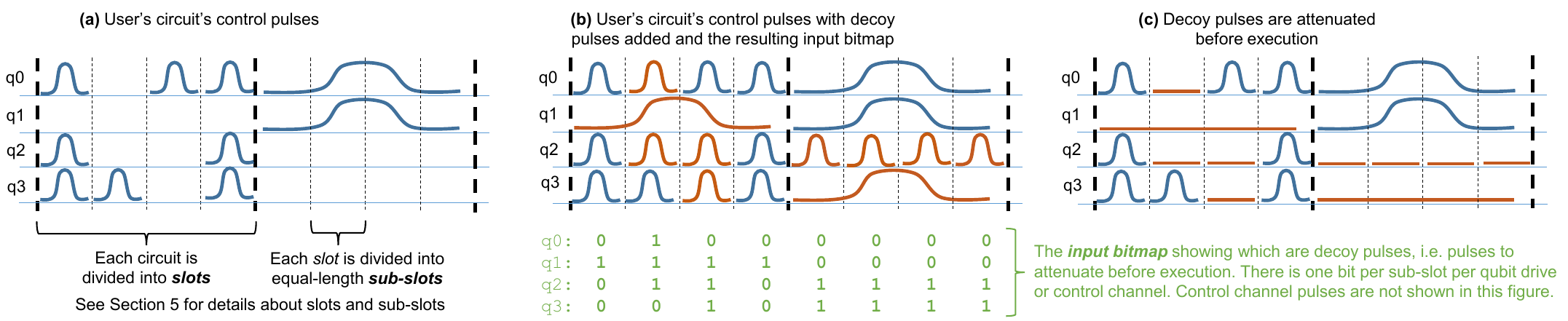}
        \caption{\small 
        Example illustration of how control pulses are obfuscated, the pulses shown are not real control pulses but their simplified graphical representation.  (a) Control pulses correspond to the input circuit of the user. (b) Control pulses  with decoy pulses added, along with the input bitmap, specifying which are real control pulses and which are decoy pulses to be attenuated. (c) Original control pulses are executed on the quantum computer, and after the attenuation of decoy pulses, attenuation is done based on the information from the input bitmap. The time is quantized into {\em sub-slots} of $160$dt. Each pulse takes an integer multiple of $160$dt, allowing for $1$ a bit in the input bitmap per qubit per {\em sub-slot} period to specify if the pulse (in that time period on the corresponding qubit) should be attenuated or not. The input bitmap is encrypted before being sent to the cloud provider. Note the figure is greatly simplified. For example, {\tt CX} gates are actually realized by multiple control pulses. Further, only drive channels are illustrated, while in real quantum computers, there are also control channels with associated control pulses used for {\tt CX} gates, decoy pulses, and input bitmap bits are needed for control channels as well.}
        \label{fig_design}
\end{figure*}
\section{QC-TEE Workflow}
Figure~\ref{fig_qctee_workflow} shows the workflow of cloud-based quantum computation after QC-TEE architecture is integrated into it. The main principle of operation is to mix the control pulses, which define the user's input quantum circuit, with decoy pulses. The control pulses mixed with the decoy pulses can be represented as a transpiled circuit that is sent to the cloud provider. In parallel, an (encrypted) input bitmap is generated to identify the location of the decoy control pulses within the transpiled circuit sent for execution. The input bitmap is encrypted with the public key of the backend, i.e. quantum computer, where the circuit will execute. It is also digitally signed by the user. Without access to the cryptographic keys of the backend, the cloud provider is not able to decrypt and read the input bitmap. The encrypted, signed input bitmap is sent to the cloud provider as well. The above steps are represented by steps $1$ -- $6$ in Figure~\ref{fig_qctee_workflow}. An illustration of the mixing is also shown in Figure~\ref{fig_design} (a) and an illustration of the input bitmap is shown in Figure~\ref{fig_design}~(b).
On the cloud provider's end, the honest-but-curious cloud provider can access and observe the control pulses mixed with the decoy pulses. Without access to the (encrypted) input bitmap, he or she is not able to tell which pulses are decoy pulses. Decoy pulses are identical to control pulses for {\tt X}, {\tt SX}, or {\tt CX} gates, i.e. when mixing in decoy pulses, pulses of all the {\tt X}, {\tt SX}, or {\tt CX} types are mixed in. When it is the user's circuit's turn to be executed on the backend, i.e. the target quantum computer, the cloud provider sends the encrypted input bitmap to the backend. Using our QC-TEE hardware in the backend, the input bitmap is decrypted within the trusted boundary of the dilution refrigerator. When the control RF pulses begin to arrive from the controller logic, the RF switches, which are part of the QC-TEE hardware, conditionally attenuate the decoy pulses. The time is quantized into periods of $160$dt. Each single qubit gate always takes $160$dt, while by design two-qubit gate's execution time is padded with delays to make it a multiple of $160$dt. With this approach, there is a need for one input bitmap bit per qubit drive or control channel, per time period. The above steps are represented by steps $7$ -- $9$ in Figure~\ref{fig_qctee_workflow}. An illustration of the attenuation of decoy pulses is also shown in Figure~\ref{fig_design} (c).
When the circuit finishes executing, the results of the measurements are sent back to the user. Optionally, as discussed in Section~\ref{mixing_algorithm}, there is a set of final {\tt X} gates which are randomly applied to the qubits before measurement. {\tt X} gates flip the state of a qubit, and whether such gate will be applied to a specific qubit is determined by the {\em output bitmap}. With a random bitmap, each qubit is flipped with a probability of 50\% before measurement. Therefore, for $n$-qubit computer, there are $2^n$ possible output combinations enabled by our addition of random {\tt X} gates. Note that the output is randomized for each {\em shot}, thus executing different shots of the same circuits yields different output. But for each shot, the user receives the (encrypted) output bitmap to correctly interpret the result.
A hardware TRNG is used to generate the random bits, one per qubit per shot.
These bits are encrypted and signed as the output bitmap and sent back to the user for each shot of the circuit. The above operations are represented by steps $10$ -- $12$ in Figure~\ref{fig_qctee_workflow}.
Lastly, the user receives the measurement outputs from the cloud provider. If random {\tt X} gates were applied at the end of the circuit, the user decrypts and verifies the output bitmap bits received. It then {\tt xor}s each of the measurement bits with the corresponding output bitmap bit. I.e. if an {\tt X} was randomly applied before measurement on a qubit $q_i$, as indicated by a $1$ bit in the output bitmap, then the measurement classical output will be flipped from the correct value. Consequently, the resulting classical bit $c_i$, corresponding to qubit $q_i$, received by the user has to be flipped back, which can be achieved by {\tt xor}ing the bit with the $1$ from the output bitmap. If an {\tt X} gate was not applied, then the output bitmap bit is $0$ and {\tt xor}ing $c_i$ with $0$ does not modify the bit. The above steps are represented by step $13$ in Figure~\ref{fig_qctee_workflow}. Finally, at step $14$, the user flips the output based on the received bitmap at the previous~step.
\section{Hardware Architecture of the QC-TEE}
\label{tee}
\label{hw_components}
\begin{figure}[t]
     \centering
         \includegraphics[width=1.0\linewidth,trim={0cm 0cm 0cm 0cm},clip]{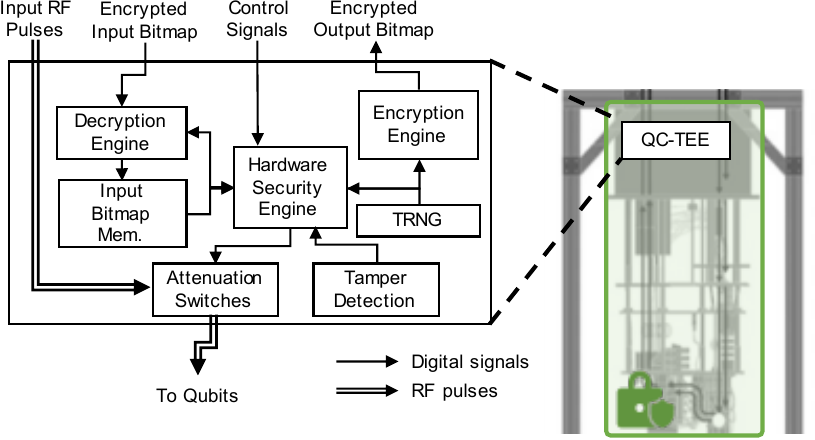}
        \caption{\small Major hardware components of the QC-TEE design added to the internals of the dilution refrigerator.}
        \label{fig_qc_tee_hw}
\end{figure}

To realize the QC-TEE, a number of hardware components need to be added to the internals of the dilution refrigerator. All components are available today and use very low power and area compared to existing quantum computer components. A block diagram of the major components and their connections is shown in Figure~\ref{fig_qc_tee_hw} and they are listed below.
\subsection{Decryption Engine and Input Bitmap Memory}
\label{sec_decryption}
A Decryption Engine is used to decrypt the encrypted input bitmap and then store the decrypted input bitmap in the Input Bitmap Memory. The decryption engine works by first using public-key cryptography to establish a shared secret key (symmetric key) and then using symmetric-key cryptography to decrypt the input bitmap itself. Both the public and private-key algorithms need to be post-quantum secure. The memory is used to store the decrypted input bitmap, which is later used to control the attenuation of the pulses.
The decryption engine consists of a Post-Quantum Cryptography Key Encapsulation Mechanism scheme (PQC-KEM) for public-key cryptography and AES-GCM \cite{NIST-AESGCM07} for symmetric-key cryptography. We propose to use a CRYSTALS-Kyber \cite{NISTPQC-R3:CRYSTALS-KYBER20} or Classic McEliece \cite{NISTPQC-R4:ClassicMcEliece22} as the PQC-KEM which will help us establish a shared secret (a key used for symmetric encryption) between the client and QC-TEE hardware. We choose these specific algorithms because they are finalists from the NIST PQC competition~\cite{nistpqc}. On the cloud provider's end, the shared secret is secured inside the fridge. These shared secrets are session based. Once the shared secret is generated, the client can use the shared secret to encrypt the input bitmap (explained in detail in Section~\ref{sec_mixer}). The AES-GCM module in the decryption engine uses the same shared secret to decrypt the input bitmap inside the dilution refrigerator of the quantum computer. In the case of the honest-but-curious cloud provider, verification of the user's input is not required, however, we suggest users to digitally sign the encrypted input bitmap, so it could be verified if needed. All the decrypted secrets are assumed to be secure from probing while inside the quantum computer's fridge, following our threat model in Section~\ref{threat_model}. The decrypted input bitmap is stored in plain text in the input bitmap memory.
\subsection{Hardware Security Engine}
Hardware Security Manager is used to send the bits of the decrypted input bitmap to the attenuation switches. Each drive and control channel is associated with one attenuation switch. In each $160$dt time period, each of the drive and control channels for all the qubits is provided with one bit: $0$ for no attenuation and $1$ for attenuation.
The hardware security engine is a hardware implementation of a state machine that controls the attenuation switches (described in Section~\ref{sec_switches}). After the decryption engine decrypts the ciphertext and stores the input bitmap in the input bitmap memory (shown in Figure~\ref{fig_qc_tee_hw}), the hardware security engine is ready and circuit execution can start. As the RF control pulses arrive, the hardware security engine reads input bitmap bits from the memory and attenuates the randomized pulses. This can be realized by triggering the hardware security engine through TTL (Transistor--Transistor logic) from the same arbitrary waveform generators that generate the RF control pulses. After the start, every $160$dt the hardware security engine outputs one control bit, per drive or control channel. The bits come from the input bitmap memory and are used to instruct the switches to attenuate (if the bit is $1$) or not (if the bit is $0$) the pulses during that time period. Note that two-qubit gates are padded with delays to make their duration a multiple of $160$dt, i.e. for two-qubit gates, multiple input bitmap bits will be used, since one bit is used per $160$dt.
In addition to that, in case of {\em randomize-output} variant of the mixer algorithm (described in Section~\ref{sec_mixer}), while executing the for the last layer of the circuit which consists of {\tt X} gates, the hardware security engine draws random bits from TRNG. The number of bits is equal to the number of qubits and attenuates if the signal on those qubits for which the generated TRNG bit is $0$ and does not attenuate if the bit is $1$. This is the opposite of input bitmap specification. However, with this specification, if the output bitmap bit is $1$, that means {\tt X} gate was applied and the state of qubit was flipped. Now, the $1$ can be also sent to the user, who can {\tt xor} the received measured bit $c_i$ with $1$ to flip it. If no {\tt X} gate was applied, the output bitmap bit is $0$, and {\tt xor} with $0$ is equivalent to no~operation.
\subsection{Attenuation Switches}
\label{sec_switches}
Attenuation Switches are used to attenuate the decoy control pulses, which were added to confuse the potential attackers, and are not actually used for computation. The simple RF switches, each requires $1$ bit of input to set if the switch should or should not attenuate the input RF signal during this time period.
Instead of adding excessive and likely unrealistic equipment inside the refrigerator, our proposed design leverages adding simple RF switches inside the refrigerator to attenuate the decoy control pulses sent to the quantum computer. 
The switches are passive elements and they do not generate any control pulses themselves. Let us consider an SPDT (single pole, double throw) RF switch, whose single pole is connected to the input line of the fridge to receive the incoming RF signal. On the output side, one of the double throws is connected to the drive line that continues to the qubit device, while the other throw is terminated by a matched impedance. This way, the SPDT RF switch can either pass or isolate the drive pulses we use to operate the quantum processor, with its on-off state controlled by a DC gate voltage. In addition, we choose GaAs-based RF switches (such as CMD196C3~\cite{cmd196c3}) that can function at cryogenic temperatures, with a wide frequency range (DC-18GHz), low insertion loss (1.5dB), high isolation (46dB), and fast switching (2.5ns). Other switches can also be used.%
\footnote{Example attenuation switches: \url{https://www.digikey.com/en/products/detail/analog-devices-inc/ADRF5040-EVALZ/6189741}}
The control voltage of the switches is generated, achieved by the hardware security engine and the input bitmap. 
We propose to mount the switches (power consumption of $\sim$ 1uW), the encryption and decryption engine (power consumption of 20mW), as well as the simple control state machine, all on the high-temperature stage (such as the 4K stage) of the fridge, which provides ample cooling power ($\sim$ 1W) to suppress the heating effect of the switch operations. The control logic and switches need to operate at a rate  between $1$ and $200$ MHz, which can be easily achieved. This is due to the fact that the single qubit gates in IBM machines each currently take $160$dt, where $1$dt $=0.222$ns, and thus each gate has to be attenuated (or not to be attenuated) at a rate of $35.5$ns ($=160 * 0.222$ns) or equivalently at a frequency of $28.5$MHz. Due to the non-idealities of the on-state of the switch (such as the insertion loss or impedance mismatch), recalibrations of the phase and amplitudes of the single- and two-qubit gate pulses are required in order to optimize the gate fatalities.
\subsection{TRNG}
\label{sec_trng}
We use the TRNG to conditionally flip the qubits before measurements when {\em randomize-output} option is enabled.
The TRNG controls which {\tt X} gates are applied, and this information is also stored in the (encrypted) output bitmap sent to the user. The TRNG is a random number generator that generates random bits based on a physical phenomenon, such as the one described in \cite{Majzoobi2011FPGABasedTR}. 
\subsection{Encryption Engine}
\label{sec_encryption} 
Encryption Engine is used if qubit flipping is enabled. The engine encrypts the output bitmap, so that the user can correctly interpret the qubit measurements. The engine uses symmetric-key encryption to encrypt the swapping bits with a randomly-generated key and public key encryption to encrypt the symmetric key with the user's public key.
The encryption engine can be a hardware implementation of an AES-GCM ~\cite{8050315}. It uses the same shared secret established as part of the decryption engine (described in Section~\ref{sec_decryption}) as a symmetric key to encrypt the TRNG output, i.e. the output bitmap, if {\em randomize-output} options is used. The client can later decrypt the encrypted output bitmap and conditionally apply logical NOT operations wherever the output bitmap indicates the corresponding bit had been flipped before measurement.
\subsection{Tamper Detection Engine}
Our work assumes the dilution refrigerator forms the trust boundary. Opening and manipulating the refrigerator can be easily detected since it is a manual process accompanied by drastic temperature and pressure changes. Qubit state will naturally be destroyed by any invasive change to the environment. Although we consider an honest-but-curious cloud provider, and we do not consider invasive attacks, we add a tamper detection engine to protect (i.e. erase) any classical information within the QC-TEE. Using a battery-based tamper detection engine, similar to ones used in classical hardware security modules (HSMs), when a temperature or pressure change is detected, all the memories within the QC-TEE hardware are erased. This erases any plaintext information about the input or output bitmaps stored in memories, and thus the attacker does not learn which are the decoy pulses or which output qubits were=flipped. Following our threat model, we assume the tamper detection engine and other QC-TEE hardware are trusted and free of bugs or trojans.
\section{Obfuscation \& De-Obfuscation}
\label{mixing_algorithm}
In this section, we describe the algorithms for how control pulses are obfuscated with the addition of decoy pulses and how later the decoy pulses are removed. 
\subsection{Obfuscator (in Software)}
\label{sec_mixer}
\begin{algorithm}[t]
\scriptsize
\begin{verbatim}
Input:  Transpiled Quantum Circuit (QC_IN), Backend 
        information (B_INFO), Obfuscation Level (OL)
Output: Randomized Transpiled Quantum Circuit (QC_OUT), 
        Input Bitmap (IB[m][n]) is used as input to the 
        quantum computer backend
        
Step 1: Split Input Circuit into Slots (Sec. 5.1.1)
        1a. Scan B_INFO and identify possible CNOT couplings 
            and identify the longest CNOT duration 
            (MAX_CNOT_DUR) and single qubit gate delay 
            (SQ_DUR)
        1b. Round MAX_CNOT_DUR up to the next even multiple of 
            single-qubit gate delay, save as CEIL_MAX_CNOT_DUR
        1c. Scan QC_IN and place a barrier before and after 
            each CNOT gate to separate single-qubit gates and 
            CNOT gates into separate slots
        1d. Insert delays in each slot:
            if (Slot_CX)
                Insert delay to pad duration of the slot to 
                CEIL_MAX_CNOT_DUR
            else if (Slot_SQ) then
                if (Obfuscation Level = quarter-delay)
                    Insert delays to pad duration of the 
                    slot to CEIL_MAX_CNOT_DUR/4
                else if (Obfuscation Level == half-delay)
                    Insert delays to pad duration of the 
                    slot to CEIL_MAX_CNOT_DUR/2
                else if (Obfuscation Level = max-delay)
                    Insert delays to pad duration of the 
                    slot to CEIL_MAX_CNOT_DUR
            
Step 2: Insert Decoy Gates into Sub-Slots (Sec. 5.1.2)
        2a. Replace delays with decoy gates:
            if (Slot_CX)
                for each coupling without a CX gate randomly
                insert CX gate, or a random mix of X and 
                SX gates totaling CEIL_MAX_CNOT_DUR; mark the 
                position of the decoy gates in Input Bitmap (IB)
            if (Slot_SQ)
                replace each delay with a random mix of X 
                and SX gates; mark the position of the decoy 
                gates in Input Bitmap
Step 3: Generate Input Bitmap (Sec. 5.1.3)
        3a. Capture information from Steps 1 and Step 2 
            for all drive and control channels C[1] to C[m] 
            in IB[1][1] to IB[m][n] (where m is the number 
            of drive and contorl channels, n is the number 
            of sub-slots on each qubit) and generate the IB:
            
            for (i from C[1] to C[m])
                for (j from IB[i][1] to IB[i][n])
                    if (gate == decoy gate)
                        B[i][j] = 1 
                    else 
                        B[i][j] = 0
                        
Step 4: Add X Gate for Randomized Output (Sec. 5.1.4)
        4a. if (randomize-output == True) then
            Append one X gate on each qubit drive channel 
            at the end of the circuit; this X gate will be 
            either executed or randomly attenuated by 
            the QC-TEE hardware
\end{verbatim}
\caption{\small Decoy Pulse Insertion Algorithm. \textit{Slot$_{CX}$} is a slot that consists of at least one {\tt CNOT} gate. \textit{Slot$_{SQ}$} is a slot that consists of only single-qubit gates.}
\label{alg_mixer}
\end{algorithm}
Our algorithm takes any transpiled quantum circuit as input, including circuits with custom gates, and performs a series of operations at the gate level to generate an obfuscated output circuit. The algorithmic description of the steps taken to insert the decoy control pulses is shown in Algorithm~\ref{alg_mixer}.
\subsubsection{Split Input Circuit into Slots}
Our algorithm takes the backend properties, such as possible {\tt CNOT} coupling and native gate duration, as input. From this, it determines the {\tt CNOT} coupling with the longest duration, as well as the duration of single-qubit gates. The algorithm then takes the input circuit already transpiled for the target backend and adds barriers%
\footnote{\url{https://qiskit.org/documentation/stubs/qiskit.circuit.library.Barrier.html}}
before and after {\tt CNOT} gates to separate portions of the circuit with single qubit gates from ones with {\tt CNOT} gates. Each barrier thus determines a start of a {\em slot}:
\begin{itemize}[noitemsep]
    \item \textit{Slot$_{CX}$}: is a slot that consists of at least one {\tt CNOT} gate and optionally any single qubit gates on the other qubits where the {\tt CNOT} gate is not connected. The duration of \textit{Slot$_{CX}$} slots is determined by choosing the longest {\tt CNOT} gate duration out of all possible couplings on the backend, and then rounding it up to be an even integer multiple of the single-gate duration. Having set the \textit{Slot$_{CX}$} duration this way, all \textit{Slot$_{CX}$} slots are checked and delays are added so that each \textit{Slot$_{CX}$} slot has the same length.
    \item \textit{Slot$_{SQ}$}: is a slot that consists of only single qubit gates and optionally any delays if required. The duration of these slots is determined at compile time and chosen based on the target security level. Currently, three duration periods are suggested: quarter-delay, half-delay and max-delay. Quarter-delay means the duration of \textit{Slot$_{SQ}$} is a quarter of the duration of \textit{Slot$_{CX}$}. Half-delay means the duration of \textit{Slot$_{SQ}$} is half of the duration of \textit{Slot$_{CX}$}, recall \textit{Slot$_{CX}$} slots are set to be of duration this is even multiple of single-qubit gate duration. Max-delay means the duration of \textit{Slot$_{SQ}$}is equal to duration of \textit{Slot$_{CX}$}.
\end{itemize}
\noindent The output of this step is a circuit that consists of multiple slots. The number of slots is determined by the depth of the circuit in terms of {\tt CNOT} gates: whichever qubit has the most {\tt CNOT} gates on it, which determines how many slots the circuit will be divided into. In our experience it is unlikely, but possible, that the number of single-qubit gates between any two {\tt CNOT} gates is more than fit in a duration of \textit{Slot$_{CX}$}. In such corner case, there will be two (or more) \textit{Slot$_{SQ}$} between consecutive \textit{Slot$_{CX}$}. In the usual case, however, \textit{Slot$_{CX}$} and \textit{Slot$_{SQ}$} occur in alternating order.
\subsubsection{Insert Decoy Gates into Sub-Slots} \label{sec_insert_decoy_gate}
The slotted input circuit is next updated by inserting decoy gates. Each slot is divided into equal-length sub-slots. The length of the sub-slot is equal to the duration of a single-qubit gate, e.g., $160$dt on IBM quantum computers today. Figure~\ref{fig_design} (a) earlier in the paper shows the input circuit with the slots and sub-slots. The sub-slots which are not occupied by a gate are filled with the decoy gates, i.e. control pulses.
For each \textit{Slot$_{CX}$} in the circuit, on the qubits that do not consist of a {\tt CNOT} gate in that slot, we either add a {\tt CNOT} gate if {\tt CNOT} coupling of the backend allows, or we add a random mix of {\tt X} and {\tt SX} gates. Figure~\ref{fig_design} (b) earlier in the paper shows (in red color) the addition of a decoy {\tt CNOT} gate next to user's {\tt CNOT} gate in a \textit{Slot$_{CX}$}.
For each \textit{Slot$_{SQ}$}, in addition to the existing single-qubit gates from the original circuit, a random mix of {\tt X} and {\tt SX} gates are added in each empty sub-slot. Figure~\ref{fig_design} (b) earlier in the paper shows (in red color) the addition of decoy single-qubit gates in a \textit{Slot$_{SQ}$}. The {\tt Rz} gates from the original circuit are added to the slots but are not counted towards a sub-slot, because the {\tt Rz} gates are virtual.
\subsubsection{Generate Input Bitmap}
While constructing the quantum circuit with the added decoy gates, we keep note of each random single-qubit gate that is inserted. In each slot, for each qubit, we divide the slot into sub-slots, each sub-slot is equal in duration to a single-qubit gate. Since one single-qubit gate fits in one sub-slot, we generate one binary bit per qubit per sub-slot to note if this is an actual gate, i.e. control pulse, or if this is a decoy pulse. The bit value $1$ represents the algorithm has inserted a decoy gate as part of obfuscation in the corresponding sub-slot -- this gate, i.e. pulse, will have to be later attenuated before it reaches the quantum computer's qubits and thus it does not actually perform any operation. Meanwhile, the bit value $0$ represents the gates that are part of the original input circuit from the original input and these should be executed, i.e. not attenuated. These bits are stored as the {\em input bitmap}. An example of the {\em input bitmap} was shown in Figure~\ref{fig_design} (b) earlier in the paper.
\subsubsection{Add {\tt X} Gate for Randomized Output} \label{sec_rand_out_sw}
In addition to the decoy gates added in the Insert Decoy Gates into Sub-Slots step (described in Section~\ref{sec_insert_decoy_gate}), an additional layer of {\tt X} gates can be added at the end of the circuit to randomize the output (shown in Step 4 of Algorithm~\ref{alg_mixer}). The randomizing of the output happens inside the dilution refrigerator on the server end. This process is described in Section~\ref{sec_outputmask}.
\subsubsection{Remove Barriers and Generate Final Circuit}
Finally, the barriers are removed from the quantum circuit while the inserted gates remain. The resulting circuit is a valid circuit that can be executed on the target backend. However, unless the decoy gates are removed, i.e. attenuated, the result of the computation will be just random. It is the purpose of the QC-TEE hardware to attenuate the decoy gates, i.e. pulses, based on the {\em input bitmap} information before these pulses actually reach qubits.
\subsubsection{Transmitting Control Pulses and Input Bitmap to Quantum Computer Backend}
When our algorithm finishes, the outputs are the transpiled and obfuscated quantum circuit, and the {\em input bitmap} indicating which gates, i.e. control pulses, in the obfuscated quantum circuit should be executed and which are to be attenuated. 
After the algorithm finishes, the {\em input bitmap} has to be encrypted and signed by using quantum-safe cryptographic algorithms before transmission to the cloud provider. Our architecture uses a key establishment mechanism (KEM) so that the user can use the public key of the quantum computer backend where the QC-TEE is located to establish a shared symmetric key for use with a standard encryption algorithm such as AES-GCM. The cloud provider has no access to the quantum computer backend's private key, so they are not able to obtain the symmetric key and decrypt the {\em input bitmap}. The obfuscated quantum circuit itself can be sent in plain text.%

\subsection{Decoy to Identity Gate Conversion (in Software)}
The output of our obfuscation scheme is a transpiled circuit which includes a number of decoy pulses. During execution, these pulses will be attenuated in hardware. The attenuation, as we discuss later in the evaluation, may not be perfect and imperfectly attenuated decoy pulses will contribute to degraded fidelity. However, we observe that during insertion of decoy pulses, there will be one or many sequences of two {\tt X} decoy gates or four {\tt SX} decoy gates that are inserted. Two {\tt X} gates in sequence or four {\tt SX} gates in sequence each form an identity~gate.
Our insight is that instead of proceeding to attenuate the sequences of two {\tt X} decoy gates or four {\tt SX} decoy gates, we can allow them to be actually executed. As identity gates, they have no effect on the circuit.
To implement the {\em decoy to identity gate conversion} we add a simple pass in software, which scans the circuit and the input bitmap. When a sequence of two {\tt X} decoy gates or four {\tt SX} decoy gates is encountered, the corresponding bits in the input bitmap are switched from $1$ (attenuate) to $0$ (don't attenuate). No further circuit modification is needed, nor is an additional transpilation step need. This makes the decoy to identity gate conversion very fast. Note that from the cloud providers' perspective, they still see all the decoy and non-decoy pulses and cannot distinguish which are real pulses, which are decoy pluses, and which are decoy pulses that were converted to identity operations. 
\subsection{De-Obfuscator (in Hardware)}
\label{sec_demixer}
The QC-TEE hardware and state machines effectively implement the inverse of the obfuscation.
The series of operations invoked by the QC-TEE hardware are as follows: We first send the ciphertext (the encrypted {\em input bitmap}) to the decryption engine (described in Section~\ref{sec_decryption}) inside the fridge. The AES-GCM module which is part of the decryption engine decrypts the ciphertext and re-generates the input bitmap containing information about which control pulses in which sub-slots should be attenuated. As shown in Figure~\ref{fig_qc_tee_hw}, this input bitmap is stored in the input bitmap memory. After this, the hardware security manager and the signal/pulse generator are notified to start generating the RF signals involved in the quantum circuit (including the decoy pulses). The attenuation switches inside the fridge (shown in Figure~\ref{fig_qc_tee_hw}) filter the unwanted random gates based on the control signals generated by the hardware security manager. We propose to achieve the synchronization between the hardware security manager and signal/pulse generator using handshake signals. Since the original quantum circuit is only de-obfuscated inside the fridge (which we consider a secure enclave), we achieve the required secrecy between client and~server. 
\subsection{Randomize Output (in Hardware)}
\label{sec_outputmask}
After the insertion of decoy pulses, the cloud provider is not easily able to recover what is the input circuit. However, by observing the generated output, a cloud provider could guess the circuit that was executed, or the generated output may be a secret value that should be protected. For this reason, we add the extra layer of {\tt X} gates if the obfuscation level is set to randomize output. The {\tt X} gates are inserted at the end of the circuit, and during execution, the control logic generates corresponding {\tt X} control pulses. However, when these {\tt X} control pulses at the end of the circuit (before the measurement gate) reach the quantum computer, QC-TEE hardware will randomly attenuate them. In software, only {\tt X} gates are inserted. The decision to attenuate them is performed randomly at runtime on the QC-TEE hardware by using the TRNG. Recall that the random bits used to specify whether to attenuate or not the final {\tt X} gates are also encrypted and sent back to the user so he or she can know how to interpret the outputs.
On receiving these encrypted output bitmap bits, the client decrypts them and then uses this information to perform logical {\tt NOT} gates on the output to recover the correct~output.

\section{Evaluation Setup}
\label{evaluation_setup}
\begin{figure}[t]
  \captionsetup[subfigure]{justification=centering}
  \centering
  \begin{subfigure}[t]{0.25\textwidth}
    \centering
    \includegraphics[width=0.46\textwidth]{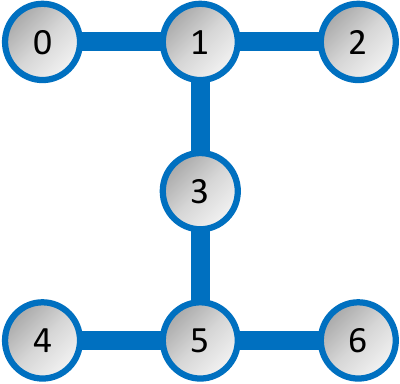}
    \caption{ibm\_perth (Falcon r5.11H)\\Basis Gates: \{CX, I, RZ, SX, X\}}
  \end{subfigure}%
  \caption{\small Device topology of $ibm\_perth$ quantum computer from IBM Quantum used in the evaluation. {\tt RZ, SX, X} are the basis single-qubit gates, {\tt CX} is the two-qubit gate and {\tt I} is the identity or delay~gate.
  }
  \label{fig_quantum_computers_used}
\end{figure}
\textbf{Quantum Computers and Simulators Used:}
To evaluate the fidelity overhead of QC-TEE hardware additions, we use a real 7-qubit IBM machine (backend), {\tt ibm\_perth}, currently available to researchers and the public through the IBM Quantum cloud-based service~\cite{ibmquantum}.
Figure~\ref{fig_quantum_computers_used} shows the topology of the quantum computer device used in this work.  We also deploy the {\tt Aer\_Simulator} and we use an imported noise model from other different real machines. 
For the {\tt Aer\_Simulator}, we run $1000$ times for each configuration per benchmark. For real machines, we are limited to one run due to our limited access to the machines.
\begin{table*}[t]
\caption{\small Average depth increase factor for the QASMBench benchmarks for IBM Perth. The average depth increase by factor is reported in each of the three Obfuscation Levels: {\em quarter-delay}, {\em half-delay} and {\em max-delay}, each of them augmented with the {\em randomize-output} option; the depth increase is computed relative to the unmodified benchmark, i.e. the {\em baseline}~configuration.}
\label{table_factor}
\small
\centering
\begin{tabular}{l|c|c|c|c|c|c}
\textbf{Obfuscation Level:}  & \multicolumn{2}{|c|}{\textbf{quarter-delay}} & \multicolumn{2}{|c|}{\textbf{half-delay}} & \multicolumn{2}{|c}{\textbf{max-delay}} \\
& \textbf{w/o rand.-out.} & \textbf{w/ rand.-out.} & \textbf{w/o rand.-out.} & \textbf{w/ rand.-out.} & \textbf{w/o rand.-out.} & \textbf{w/ rand.-out.} \\
\hline
\textbf{Real IBM Perth ($QV=32$)} & $8.25$x & $8.30$x & $9.45$x & $9.50$x & $11.51$x & $11.55$x \\
\end{tabular}
\end{table*}
\begin{table*}[t]
\caption{\small Average Variational Distance ($VD$) for the QASMBench benchmarks for machines with different Quantum Volume ($QV$). The average $VD$ is reported in each of the three Obfuscation Levels: {\em quarter-delay}, {\em half-delay} and {\em max-delay}, each of them augmented with the {\em randomize-output} option; the $VD$ is computed relative to the unmodified benchmark, i.e. the {\em baseline}~configuration. The data is for ideal RF~switches.}
\label{table_qv}
\small
\centering
\begin{tabular}{l|c|c|c|c|c|c}
\textbf{Obfuscation Level:}  & \multicolumn{2}{|c|}{\textbf{quarter-delay}} & \multicolumn{2}{|c|}{\textbf{half-delay}} & \multicolumn{2}{|c}{\textbf{max-delay}} \\
& \textbf{w/o rand.-out.} & \textbf{w/ rand.-out.} & \textbf{w/o rand.-out.} & \textbf{w/ rand.-out.} & \textbf{w/o rand.-out.} & \textbf{w/ rand.-out.} \\
\hline
\textbf{Real IBM Perth ($QV=32$)} & $0.1622$ & $0.1622$ & $0.1647$ & $0.1827$ & $0.2099$ & $0.2453$ \\
\hline
\textbf{Sim. IBM Perth ($QV=32$)} & $0.0981$ & $0.1024$ & $0.1289$ & $0.1302$ & $0.1461$  & $0.1481$ \\
\hline
\textbf{Sim. IBM Mumbai ($QV=64$)} & $0.1014$ & $0.1045$ & $0.1379$ & $0.1395$ & $0.1653$ & $0.1705$ \\
\hline
\textbf{Sim. IBM Cairo ($QV=128$)} & $0.0689$ & $0.0702$ & $0.0806$ & $0.0856$ & $0.1286$ & $0.1299$ \\
\end{tabular}
\end{table*}
\begin{table*}[t!]
\caption{\small Average Variational Distance ($VD$) for the QASMBench benchmarks when considering perfect RF switches, when emulating imperfect RF switches which attenuate the decoy pulses to $0.01$\% amplitude and when we convert decoy gates to identity gates. Data is from experiments on IBM Perth.
}
\label{table_imperfect_switches}
\small
\centering
\begin{tabular}{p{4cm}|c|c|c|c|c|c}
\textbf{Obfuscation Level:}  & \multicolumn{2}{|c|}{\textbf{quarter-delay}} & \multicolumn{2}{|c|}{\textbf{half-delay}} & \multicolumn{2}{|c}{\textbf{max-delay}} \\
& \textbf{w/o rand.-out.} & \textbf{w/ rand.-out.} & \textbf{w/o rand.-out.} & \textbf{w/ rand.-out.} & \textbf{w/o rand.-out.} & \textbf{w/ rand.-out.} \\
\textbf{Avg. VD} (perfect RF switches) & $0.1622$ & $0.1622$ & \textbf{$0.1647$} & $0.1827$  & \textbf{$0.2099$} &  \textbf{$0.2453$}\\
\hline
\hline
\textbf{Avg. VD} (imperfect RF switches) & $0.2248$ & $0.2100$ & \textbf{$0.2143$} & $0.2143$ & \textbf{$0.2652$} &  \textbf{$0.2691$}\\
\hline
\textbf{Avg. VD} (imperfect RF switches w/ identity gate conversion) & $0.1548$  & $0.1556$ & $0.1602$  & $0.1613$ & $0.1998$ &  $0.2188$\\
\end{tabular}
\end{table*}
\textbf{Benchmarks Used:}
QASMBench Benchmark Suite version 1.4~\cite{qasmbenchs} is used in our work to analyze the impact of the QC-TEE design on different circuits. We transpile each benchmark for the target backend (either for the real machine or for the simulator with imported noise models). The transpiled code is then used as input to the obfuscation algorithm, which is described in Section~\ref{mixing_algorithm}. QASMBench includes three categories of benchmarks of different sizes and we use the first category, which refers to the small-scale quantum circuits that use 2 to 10 qubits. Because our access is limited to a real quantum computer with $7$ qubits, we only select the benchmarks that can be run on this machine. We keep the number of shots constant at $8192$. For the simulator, we use the same set of benchmarks, i.e., small-scale.
\textbf{Obfuscation Levels Evaluated:}
We evaluated five different configurations: {\em baseline}, {\em quarter-delay}, {\em half-delay}, {\em max-delay} and {\em randomize-output}. The {\em baseline} is simply the benchmark transpiled for the target backend without any of our modifications. The {\em quarter-delay} means the duration of \textit{Slot$_{SQ}$} is a quarter of the duration of \textit{Slot$_{CX}$} and the {\em half-delay} means the duration of \textit{Slot$_{SQ}$} is half of the duration of \textit{Slot$_{CX}$}, recall \textit{Slot$_{CX}$} slots are set to be of duration, which is even multiple of single-qubit gate duration. The {\em max-delay} means the duration of \textit{Slot$_{SQ}$} is equal to the duration of \textit{Slot$_{CX}$}. The {\em randomize-output} includes an additional layer of {\tt X} gates before measurement.
\textbf{Variational Distance:}
We measure the variational distance between each of the three configurations {\em quarter-delay}, {\em half-delay}, {\em max-delay}, all of them with and without the {\em randomize-output} option, and the {\em baseline} in order to define it as a metric to examine how our design affected the output probability of the benchmark circuits. Informally, the variational distance of two output probability distributions is the measure of how one probability distribution is different from the other. In general, the total variation distance between P and Q is defined as:
\[ \delta(P, Q) = \frac{1}{2} \sum|P - Q| \]

\section{Evaluation}
In this section, we present the evaluation of our design. Since we are not able to modify IBM Quantum computers to insert the RF switches and the QC-TEE logic, we focus on the evaluation of the expected impact of the changes on the fidelity of the circuits.
\subsection{Impact of Increased Circuit Duration}
\label{circuit_fidelity}
Our obfuscation technique works by adding decoy pulses and then attenuating them before actual execution. Once the decoy pulses are attenuated, they effectively become delays. The overhead of the added delays is shown in Table~\ref{table_factor}. 
For the simulator, we measured the average of $1000$ experiments for each scenario and each benchmark and we also present the error bars. In the presence of ideal RF switches, the circuit that is executed is the same as provided by the user, but with extra delays (where the decoy pulses were). The first row of Table~\ref{table_imperfect_switches} shows the average Variational Distance ($VD$) for the different obfuscation levels over the set of QASMBench benchmarks, assuming perfect RF switches. We observe, as expected, that with more delays, the variational distance increases. Thus, with ideal RF switches, the fidelity decrease, i.e. increase of the $VD$, can be attributed to the decoherence characteristics of the underlying hardware. However, assuming Schoelkopf's law, which states that roughly every three years the quantum decoherence has been improved by a factor of 10, we expect future machines to be getting better, resulting in QC-TEE delays having less and less impact on the fidelity of the circuits.
\subsection{Sensitivity to Quantum Volume and Noise}
We explored if and how Quantum Volume and Noise model may impact the Variational Distance. Table~\ref{table_qv} reports the average $VD$ over the set of the QASMBench benchmarks.
In Table~\ref{table_qv} we explore machines with different Quantum Volumes. We observe a lower average $VD$ for higher quantum volume machines. We used the imported noise model of IBM Mumbai and IBM Cairo, since we have no access to real machines with quantum volumes higher than $32$. The table shows IBM Perth for both real machines and simulation with the imported noise model. We conclude that machines with bigger quantum volumes will benefit QC-TEE.
\subsection{Evaluation of Imperfect RF Switches}
The key component of the QC-TEE is the RF switches. In order to understand how imperfections of the switches affect the fidelity, we performed tests where custom pulses were used to emulate an imperfect RF switch. Specifically, we developed custom gates that have the same pulse shape as {\tt X}, {\tt SX}, and {\tt CX} gates, but they have only $0.01$\% amplitude to simulate imperfect attenuation by RF switches. Next, rather than replacing the decoy gates with delays (which represents ideal RF switches with 100\% attenuation), we replaced the decoy pulses with the $0.01$\% amplitude custom gates. These gates were then executed on a real IBM backend.
Table~\ref{table_imperfect_switches} shows the results. The baseline in this case is the circuits with delays, so we can understand how imperfections in the attenuation affect Variational Distance.
\subsection{Evaluation of Fidelity}
Although added decoy gates degrade the fidelity of the circuits, in all tested algorithms, outputs are correct if the variational distance is less than $0.35$. The RF switches CMD196C can be expected to attenuate the decoy pulses to 0.01\% amplitude, the second row of Table~\ref{table_imperfect_switches} shows average variational distances over the tested benchmarks assuming such switches. With better (ideal) RF switches, the variational distance becomes less, top row of Table~\ref{table_imperfect_switches}).  On average, even with today's imperfect RF switches, benchmarks will give correct results.
In case of all circuits, but especially for circuits that may give wrong output when decoy gates cause too much noise, our {\em decoy to identity gate conversion} can be applied. In all cases, the conversion will improve the fidelity, decreasing the variational distance as shown in the last row of Table~\ref{table_imperfect_switches}.
\subsection{Evaluation of Power Requirements}
\label{sec_power_eval}
While deciding on the cryptographic algorithms used in QC-TEE, in addition to security, it is also important to consider the power consumption of these algorithms since the hardware running these algorithms would be inside the dilution refrigerator.
For the public key cryptographic algorithm, we note that the existing work \cite{cryptoeprint:2023/506} shows that for CRYSTALS-Kyber with the highest security level, the Key Generation, Encapsulation, and Decapsulation consume 157.2  mW, 160.4mW, and 162.0mW, respectively on a Xilinx Artix 7 FPGA. And for symmetric key cryptography, AES-GCM~\cite{8050315} consumes 19.52mW  for encryption/decryption on Altera Cyclone V FPGA. We note that while we were writing this work, no prior work on running cryptographic algorithms in cryogenic temperatures was documented in the literature to the best of our knowledge. However, we note that other applications, such as \cite{10.1063/1.4939094}, have been successfully tested in cryogenic temperatures ($4K$). In addition to that, the authors also noted that the overall operating frequency improved while running the FPGA application at cryogenic temperature.
\subsection{Evaluation of Physical Overhead}
\label{sec_overhead_eval}
\begin{table*}[t]
\caption{\small Estimated power consumption and estimated volume overhead due to QC-TEE hardware additions inside the dilution refrigerator.
}
\label{tab:estimated_power}
\small
\centering
\begin{tabular}{cc|cc|c||c||c}
\textbf{Qubits} &\multicolumn{1}{c}{+}& \textbf{Couplings} &\multicolumn{1}{c}{=}& \textbf{RF Switches} & \textbf{Power Overhead (mW)} & \multicolumn{1}{l}{\textbf{Volume Overhead (\%)}} \\
\hline
$27$ (Falcon)      && $28$   && $55$   & $181.57$   & 0.00003\% \\ \hline
$65$ (Hummingbird) && $72$   && $137$  & $181.65$   & 0.00009\% \\ \hline
$127$ (Eagle)      && $144$  && $271$  & $181.79$   & 0.00018\% \\ \hline
$433$ (Osprey)     && $504$  && $937$  & $182.45$   & 0.00064\% \\ \hline
$1121$ (Condor)    && $1186$ && $2242$ & $183.76$   & 0.00157\% \\ \hline
$1386$ (Flamingo)  && $1387$ && $2773$ & $184.29$   & 0.00189\% \\
\end{tabular}
\end{table*}
The main overhead of the scheme comes in terms of power, evaluated in above in Section~\ref{sec_power_eval}, as well as physical overhead of the QC-TEE logic and RF switches. Considering XLD Blue Fors dilution refrigerator~\cite{blueforsCryogenfreeXLDsl}
the volume of the refrigerator with height of about $1481 mm$ and radius of about $460mm$ is $\pi \times 460^2 \times 1481 = 984,011,944 mm^3$. Bigger refrigerators will be needed for larger quantum computers, but taking this as a conservative estimate, Table~\ref{tab:estimated_power} how the percentage of volume is taken by the added hardware for quantum computers with different qubit size. Note that the overhead is in all cases is $0.002$\% or less.
\subsection{Evaluation of Practicality and Scalability}
\label{sec_scalability_eval}
The proposed QC-TEE can scale to handle large number of qubits. We note that publicly announced designs for large quantum computers are modular. For example, the forthcoming Flamingo processor~\cite{roadmap} is projected to support 1,386 qubits, which will be located in three 462-qubit modules or chip. One QC-TEE can be deployed per module or chip. In Table~\ref{tab:estimated_power} we show the number of RF switches and the expected power consumption of the QC-TEE logic for different size quantum computers from 27 to 1,386 -- the sizes are taken from the IBM roadmap~\cite{roadmap}. Note that larger quantum processors, such as Kookaburra, will be built from these smaller modules and would simply require a number of QC-TEE hardware proportional to the number of the quantum computing.

\section{Security Discussion}
\label{discussion}
\subsection{Security Analysis}
To evaluate the security of our proposed scheme, we compute the total number of possible circuits that the obfuscated circuit could represent. This is the attack complexity from the perspective of the honest-but-curious cloud provider. In the computation, we take into account the following aspects: 
number of qubits ($n_{qubits}$), 
number of {\tt CNOT} gate slots ($n_{Slot_{CX}}$), 
number of single qubit gate slots ($n_{Slot_{SQ}}$), 
number of gates {which fit in a slot %
$Slot_{SQ}$ on each qubit ($n_{SubSlots}$), 
number of {\tt CNOT} gates%
on different qubits in a $Slot_{CX}$ ($n_{SubCXInSlotCX}$), 
number of single qubit gates which fit on qubits not consisting {\tt CNOT} gates in $Slot_{CX}$ ($n_{SubSlotsInSlotCX}$) 
and insertion of {\tt X} gates for the randomized output option. 
The number of combinations ($Comb$) that the malicious cloud provider must try to find the correct circuit executed inside the fridge is:
\begin{multline*}
Comb = (n_{qubits} \times ({2}^{n_{SubSlots}}) \times n_{Slot_{SQ}}) \times \\ ((2 \times n_{SubCXInSlotCX}) + (n_{qubits}-2 \times {n_{SubCXInSlotCX}}) \\ \times2^{n_{SubSlotsInSlotCX}})  \times n_{Slot_{CX}} \times 2^{n_{qubits}}
\end{multline*}
Considering the benchmarks we have evaluated, the complexity depends on the benchmark and the number of qubits and gates that the benchmark uses. For example, for {\em qaoa} benchmark, which uses $6$ qubits, $270$ gates, $54$ {\tt CNOT} gates (before transpilation), the attack complexity for quarter-delay is $2^{24.32}$, for half-delay is $2^{29.32}$, and for max-delay is $2^{38.32}$. For circuits with more qubits or gates, the complexity increases exponentially in the number of qubits. For example, for the large {\em adder} benchmark, which uses $433$ qubits, $6769$ gates, $3120$ {\tt CNOT} gates, the attack complexity is on the order of $2^{470}$. Regardless of circuit size, users can always add dummy qubits (and thus decoy gates) to increase the complexity~further.
\subsection{Timing Side-Channel Attacks on User Circuits}
Our design focuses on honest-but-curious cases where the cloud provider wants to learn what circuits the user is executing, without any prior knowledge. However, if the cloud provider has additional knowledge, such as a list of possible circuits and their duration in time, they can leverage this information plus the real execution timing to perform the user circuit (UC) identification attack.
In order to protect from such timing side-channel attacks, the next step is to make the number of slots to be variable and not directly depend on the circuits. For example, a random number of extra \textit{Slot$_{CX}$} and \textit{Slot$_{SQ}$} can be added into the circuits, where each such slot is full of decoy control pulses. Therefore, circuit duration can be elongated and thus may be prevented from timing attacks.
\subsection{Power Side-Channel Attacks on QC-TEE}
There can also be power side-channel attacks on quantum computers, which have been proposed in concurrent work in~\cite{xu2023exploration}. However, the authors only assume the power can be measured from the drive equipment. If the on and off states of switches have an influence on the power and can be measured by malicious providers, then the providers may be able to recover the real circuits. This may require additional defense from the hardware side. Our work assumes the fridge forms a trust boundary and power attacks on equipment inside the fridge are out of scope.

\section{Related Work}
\label{related_work}
To help protect from the untrusted quantum computer cloud providers, delegated quantum computation (DQC) and blind quantum computation (BQC) have been proposed~\cite{universal_blind, aharonov2008interactive, morimae2011ground, Dunjko_20121, Morimae_2012, Morimae_2013, Fitzsimons_20171, Morimae_20121, PhysRevA.87.060301, morimae2013composable, Giovannetti_2013, Mantri_2013, morimae2014verification}. Most of the work remains theoretical, due to the fact that DQC and BQC require a local, trusted quantum computer, as well as, quantum networking used to connect the remote and local quantum computers, which is not available today.
Meanwhile, QC-TEE can be realized today with minor additions of commodity hardware such as the RF switches.
Approaches which do not use blind quantum computation include concurrent work~\cite{patel2023toward} that at compile time adds pairs of {\tt RX} gates into the circuit to obfuscate it to the cloud provider. The work however does not leverage any trusted hardware, and all the inserted gates are actually executed, severely limiting the amount of possible obfuscation compared to our work.
Idea to use {\tt X} gates before measurement to reduce measurement errors has been proposed in~\cite{tannu2019mitigating}. We also leverage the idea of inserting {\tt X} gates, but for the purpose of protecting output from the cloud provider, only a user who can access the encrypted output bitmap can recover the correct output.

\section{Conclusion}
\label{conclusion}
This work presented the first design for a trusted execution environment for quantum computers, to protect from honest-but-curious cloud providers who may try to steal or learn what quantum program the user is executing. The proposed QC-TEE architecture imposes minimal hardware changes, mostly the addition of RF switches and related control logic, to the fridge of a superconducting qubit quantum computer. To protect the circuits, they are obfuscated with decoy control pulses added during circuit transpilation by the user. The decoy pulses are removed, i.e. attenuated, by the QC-TEE hardware before they reach the qubits. We also provided an option to protect outputs with random application of {\tt X} gates, as well as we explored the conversion of decoy gates to identity gates to improve fidelity. Users of QC-TEE can select between different levels of obfuscation to trade off security for circuit fidelity.

\bibliographystyle{IEEEtranS}
\bibliography{main}

\begin{thebibliography}{10}
\providecommand{\url}[1]{#1}
\csname url@samestyle\endcsname
\providecommand{\newblock}{\relax}
\providecommand{\bibinfo}[2]{#2}
\providecommand{\BIBentrySTDinterwordspacing}{\spaceskip=0pt\relax}
\providecommand{\BIBentryALTinterwordstretchfactor}{4}
\providecommand{\BIBentryALTinterwordspacing}{\spaceskip=\fontdimen2\font plus
\BIBentryALTinterwordstretchfactor\fontdimen3\font minus
  \fontdimen4\font\relax}
\providecommand{\BIBforeignlanguage}[2]{{%
\expandafter\ifx\csname l@#1\endcsname\relax
\typeout{** WARNING: IEEEtranS.bst: No hyphenation pattern has been}%
\typeout{** loaded for the language `#1'. Using the pattern for}%
\typeout{** the default language instead.}%
\else
\language=\csname l@#1\endcsname
\fi
#2}}
\providecommand{\BIBdecl}{\relax}
\BIBdecl

\bibitem{amazonbracket}
``Amazon braket,'' \url{https://aws.amazon.com/braket/}.

\bibitem{azurequantum}
``Azure quantum,'' \url{https://azure.microsoft.com/en-us/products/quantum}.

\bibitem{cmd196c3}
``Cmd196c3,'' \url{https://www.qorvo.com/products/d/da007444}.

\bibitem{blueforsCryogenfreeXLDsl}
``{C}ryogen-free {X}{L}{D}sl dilution refrigerator measurement system -
  {B}luefors --- bluefors.com,''
  \url{https://bluefors.com/products/xldsl-dilution-refrigerator/}, [Accessed
  05-08-2023].

\bibitem{ibmquantum}
``Ibm quantum,'' \url{https://quantum-computing.ibm.com/}.

\bibitem{roadmap}
``Ibm quantum development roadmap,'' \url{https://www.ibm.com/quantum/roadmap}.

\bibitem{4000qubits}
``Ibm's target:a 4000-qubit processor by 2025,''
  \url{https://spectrum.ieee.org/ibm-quantum-computer}.

\bibitem{nistpqc}
``Nist pqc competition,''
  \url{https://csrc.nist.gov/projects/post-quantum-cryptography/round-4-submissions}.

\bibitem{qasmbenchs}
``Qasmbench,'' \url{https://github.com/pnnl/QASMBench}.

\bibitem{quantumcircuitsinc}
``Quantum circuits,'' \url{https://quantumcircuits.com/}.

\bibitem{rigetti}
``Rigetti computing,'' \url{https://www.rigetti.com/}.

\bibitem{childs}
\BIBentryALTinterwordspacing
sep 2005, \url{https://arxiv.org/pdf/quant-ph/0111046.pdf}. [Online].
  Available: \url{https://doi.org/10.26421%2Fqic5.6}
\BIBentrySTDinterwordspacing

\bibitem{aharonov2008interactive}
D.~Aharonov, M.~Ben-Or, and E.~Eban, ``Interactive proofs for quantum
  computations,'' 2008, \url{https://arxiv.org/pdf/0810.5375.pdf}.

\bibitem{NISTPQC-R4:ClassicMcEliece22}
M.~R. Albrecht, D.~J. Bernstein, T.~Chou, C.~Cid, J.~Gilcher, T.~Lange,
  V.~Maram, I.~von Maurich, R.~Misoczki, R.~Niederhagen, K.~G. Paterson,
  E.~Persichetti, C.~Peters, P.~Schwabe, N.~Sendrier, J.~Szefer, C.~J. Tjhai,
  M.~Tomlinson, and W.~Wang, ``{Classic McEliece},'' {N}ational {I}nstitute of
  {S}tandards and {T}echnology, Tech. Rep., 2022, available at
  \url{https://csrc.nist.gov/Projects/post-quantum-cryptography/round-4-submissions}.

\bibitem{universal_blind}
A.~Broadbent, J.~Fitzsimons, and E.~Kashefi, ``Universal blind quantum
  computation,'' in \emph{2009 50th Annual IEEE Symposium on Foundations of
  Computer Science}, 2009, pp. 517--526,
  \url{https://ieeexplore.ieee.org/stamp/stamp.jsp?tp=&arnumber=5438603}.

\bibitem{chow2021ibm}
J.~Chow, O.~Dial, and J.~Gambetta, ``Ibm quantum breaks the 100-qubit processor
  barrier,'' \emph{IBM Research Blog}, vol.~2, 2021.

\bibitem{10.1063/1.4939094}
\BIBentryALTinterwordspacing
I.~D. Conway~Lamb, J.~I. Colless, J.~M. Hornibrook, S.~J. Pauka, S.~J. Waddy,
  M.~K. Frechtling, and D.~J. Reilly, ``{An FPGA-based instrumentation platform
  for use at deep cryogenic temperatures},'' \emph{Review of Scientific
  Instruments}, vol.~87, no.~1, 01 2016, 014701. [Online]. Available:
  \url{https://doi.org/10.1063/1.4939094}
\BIBentrySTDinterwordspacing

\bibitem{intelsgx}
V.~Costan and S.~Devadas, ``Intel sgx explained,'' \emph{Cryptology ePrint
  Archive}, 2016.

\bibitem{cross2018ibm}
A.~Cross, ``The ibm q experience and qiskit open-source quantum computing
  software,'' in \emph{APS March meeting abstracts}, vol. 2018, 2018, pp.
  L58--003.

\bibitem{Dunjko_2014}
\BIBentryALTinterwordspacing
V.~Dunjko, J.~F. Fitzsimons, C.~Portmann, and R.~Renner, ``Composable security
  of delegated quantum computation,'' in \emph{Lecture Notes in Computer
  Science}.\hskip 1em plus 0.5em minus 0.4em\relax Springer Berlin Heidelberg,
  2014, pp. 406--425, \url{https://arxiv.org/pdf/1301.3662.pdf}. [Online].
  Available: \url{https://doi.org/10.1007%2F978-3-662-45608-8_22}
\BIBentrySTDinterwordspacing

\bibitem{Dunjko_20121}
\BIBentryALTinterwordspacing
V.~Dunjko, E.~Kashefi, and A.~Leverrier, ``Blind quantum computing with weak
  coherent pulses,'' \emph{Physical Review Letters}, vol. 108, no.~20, may
  2012, \url{https://arxiv.org/pdf/1108.5571.pdf}. [Online]. Available:
  \url{https://doi.org/10.1103%2Fphysrevlett.108.200502}
\BIBentrySTDinterwordspacing

\bibitem{NIST-AESGCM07}
M.~Dworkin, ``{Recommendation for Block Cipher Modes of Operation:
  Galois/Counter Mode (GCM) and GMAC},'' {N}ational {I}nstitute of {S}tandards
  and {T}echnology, Tech. Rep., 2007, available at
  \url{https://csrc.nist.gov/publications/detail/sp/800-38d/final}.

\bibitem{https://doi.org/10.48550/arxiv.1611.10107}
\BIBentryALTinterwordspacing
J.~F. Fitzsimons, ``Private quantum computation: An introduction to blind
  quantum computing and related protocols,'' 2016,
  \url{https://www.nature.com/articles/s41534-017-0025-3.pdf}. [Online].
  Available: \url{https://arxiv.org/abs/1611.10107}
\BIBentrySTDinterwordspacing

\bibitem{Fitzsimons_20171}
\BIBentryALTinterwordspacing
J.~F. Fitzsimons and E.~Kashefi, ``Unconditionally verifiable blind quantum
  computation,'' \emph{Physical Review A}, vol.~96, no.~1, jul 2017,
  \url{https://arxiv.org/pdf/1203.5217.pdf}. [Online]. Available:
  \url{https://doi.org/10.1103%2Fphysreva.96.012303}
\BIBentrySTDinterwordspacing

\bibitem{Giovannetti_2013}
\BIBentryALTinterwordspacing
V.~Giovannetti, L.~Maccone, T.~Morimae, and T.~G. Rudolph, ``Efficient
  universal blind quantum computation,'' \emph{Physical Review Letters}, vol.
  111, no.~23, dec 2013, \url{https://arxiv.org/pdf/1306.2724.pdf}. [Online].
  Available: \url{https://doi.org/10.1103%2Fphysrevlett.111.230501}
\BIBentrySTDinterwordspacing

\bibitem{8050315}
S.~Koteshwara, A.~Das, and K.~K. Parhi, ``Fpga implementation and comparison of
  aes-gcm and deoxys authenticated encryption schemes,'' in \emph{2017 IEEE
  International Symposium on Circuits and Systems (ISCAS)}, 2017, pp. 1--4.

\bibitem{Majzoobi2011FPGABasedTR}
M.~Majzoobi, F.~Koushanfar, and S.~Devadas, ``Fpga-based true random number
  generation using circuit metastability with adaptive feedback control,'' in
  \emph{Workshop on Cryptographic Hardware and Embedded Systems}, 2011.

\bibitem{Mantri_2013}
\BIBentryALTinterwordspacing
A.~Mantri, C.~A. P{\'{e} }rez-Delgado, and J.~F. Fitzsimons, ``Optimal blind
  quantum computation,'' \emph{Physical Review Letters}, vol. 111, no.~23, dec
  2013, \url{https://arxiv.org/pdf/1306.3677.pdf}. [Online]. Available:
  \url{https://doi.org/10.1103%2Fphysrevlett.111.230502}
\BIBentrySTDinterwordspacing

\bibitem{MauRen11}
U.~Maurer and R.~Renner, ``Abstract cryptography,'' in \emph{The Second
  Symposium on Innovations in Computer Science, ICS 2011}, B.~Chazelle,
  Ed.\hskip 1em plus 0.5em minus 0.4em\relax Tsinghua University Press, 1 2011,
  \url{https://crypto.ethz.ch/publications/files/MauRen11.pdf}.

\bibitem{mavrovouniotis2013hardware}
S.~Mavrovouniotis and M.~Ganley, ``Hardware security modules,'' in \emph{Secure
  Smart Embedded Devices, Platforms and Applications}.\hskip 1em plus 0.5em
  minus 0.4em\relax Springer, 2013, pp. 383--405.

\bibitem{Morimae_20121}
\BIBentryALTinterwordspacing
T.~Morimae, ``Continuous-variable blind quantum computation,'' \emph{Physical
  Review Letters}, vol. 109, no.~23, dec 2012,
  \url{https://arxiv.org/pdf/1208.0442.pdf}. [Online]. Available:
  \url{https://doi.org/10.1103%2Fphysrevlett.109.230502}
\BIBentrySTDinterwordspacing

\bibitem{morimae2014verification}
------, ``Verification for measurement-only blind quantum computing,'' 2014,
  \url{https://arxiv.org/pdf/1208.1495.pdf}.

\bibitem{morimae2011ground}
T.~Morimae, V.~Dunjko, and E.~Kashefi, ``Ground state blind quantum computation
  on aklt state,'' 2011, \url{https://arxiv.org/pdf/1009.3486.pdf}.

\bibitem{Morimae_2012}
\BIBentryALTinterwordspacing
T.~Morimae and K.~Fujii, ``Blind topological measurement-based quantum
  computation,'' \emph{Nature Communications}, vol.~3, no.~1, sep 2012,
  \url{https://arxiv.org/pdf/1110.5460.pdf}. [Online]. Available:
  \url{https://doi.org/10.1038%2Fncomms2043}
\BIBentrySTDinterwordspacing

\bibitem{Morimae_2013}
\BIBentryALTinterwordspacing
------, ``Blind quantum computation protocol in which alice only makes
  measurements,'' \emph{Physical Review A}, vol.~87, no.~5, may 2013,
  \url{https://arxiv.org/pdf/1201.3966.pdf}. [Online]. Available:
  \url{https://doi.org/10.1103%2Fphysreva.87.050301}
\BIBentrySTDinterwordspacing

\bibitem{morimae2013composable}
T.~Morimae and T.~Koshiba, ``Composable security of measuring-alice blind
  quantum computation,'' 2013, \url{https://arxiv.org/pdf/1306.2113.pdf}.

\bibitem{patel2023toward}
T.~Patel, D.~Silver, A.~Ranjan, H.~Gandhi, W.~Cutler, and D.~Tiwari, ``Toward
  privacy in quantum program execution on untrusted quantum cloud computing
  machines for business-sensitive quantum needs,'' 2023.

\bibitem{NISTPQC-R3:CRYSTALS-KYBER20}
P.~Schwabe, R.~Avanzi, J.~Bos, L.~Ducas, E.~Kiltz, T.~Lepoint, V.~Lyubashevsky,
  J.~M. Schanck, G.~Seiler, and D.~Stehl{\'e}, ``{CRYSTALS-KYBER},'' {N}ational
  {I}nstitute of {S}tandards and {T}echnology, Tech. Rep., 2020, available at
  \url{https://csrc.nist.gov/projects/post-quantum-cryptography/round-3-submissions}.

\bibitem{PhysRevA.87.060301}
\BIBentryALTinterwordspacing
T.~Sueki, T.~Koshiba, and T.~Morimae, ``Ancilla-driven universal blind quantum
  computation,'' \emph{Phys. Rev. A}, vol.~87, p. 060301, Jun 2013. [Online].
  Available: \url{https://link.aps.org/doi/10.1103/PhysRevA.87.060301}
\BIBentrySTDinterwordspacing

\bibitem{tannu2019mitigating}
S.~S. Tannu and M.~K. Qureshi, ``Mitigating measurement errors in quantum
  computers by exploiting state-dependent bias,'' in \emph{Proceedings of the
  52nd annual IEEE/ACM international symposium on microarchitecture}, 2019, pp.
  279--290.

\bibitem{cryptoeprint:2023/506}
\BIBentryALTinterwordspacing
G.~Tasopoulos, C.~Dimopoulos, A.~P. Fournaris, R.~K. Zhao, A.~Sakzad, and
  R.~Steinfeld, ``Energy consumption evaluation of post-quantum tls 1.3 for
  resource-constrained embedded devices,'' Cryptology ePrint Archive, Paper
  2023/506, 2023, \url{https://eprint.iacr.org/2023/506}. [Online]. Available:
  \url{https://eprint.iacr.org/2023/506}
\BIBentrySTDinterwordspacing

\bibitem{wang2022quantum}
J.~Wang, L.~Liu, S.~Lyu, Z.~Wang, M.~Zheng, F.~Lin, Z.~Chen, L.~Yin, X.~Wu, and
  C.~Ling, ``Quantum-safe cryptography: crossroads of coding theory and
  cryptography,'' \emph{Science China Information Sciences}, vol.~65, no.~1, p.
  111301, 2022.

\bibitem{xu2023exploration}
C.~Xu, F.~Erata, and J.~Szefer, ``Exploration of quantum computer power
  side-channels,'' 2023.

\end{thebibliography}
\end{document}